\documentclass[%
 reprint,
 amsmath,amssymb,
 aps,
]{revtex4-2}

\usepackage{graphicx}
\usepackage{dcolumn}
\usepackage{bm}
\usepackage{float}

\newcommand{\beq}{\begin{equation}}
\newcommand{\eeq}{\end{equation}}

\newcommand{\bu}{{\bf u}}

\newcommand{\bk}{{\bf k}}

\newcommand{\bx}{{\bf x}}
\newcommand{\br}{{\bf r}}

\newcommand{\curl}{\nabla \times}

\begin{document}

\preprint{APS/123-QED}

\title{Inducing intermittency in the inverse cascade \\ of two dimensional turbulence by a fractal forcing}

\author{George Sofiadis}
\affiliation{%
Department of Mechanical Engineering, University of West Attica, 250 Thivon \& P. Ralli Str., Egaleo, 122 44, Athens, Greece
}%
\author{Ioannis E. Sarris}%
 
\affiliation{%
Department of Mechanical Engineering, University of West Attica, 250 Thivon \& P. Ralli Str., Egaleo, 122 44, Athens, Greece
}%


\author{Alexandros Alexakis}
\email{alexakis@phys.ens.fr}
\affiliation{
Laboratoire de Physique de l’Ecole normale supérieure, ENS, Université PSL, CNRS, Sorbonne Université, Université de Paris, F-75005 Paris, France
}%

\date{\today}

\begin{abstract}
We demonstrate that like in the forward cascade of three dimensional turbulence that displays intermittency (lack of self-similarity) due to the concentration of energy dissipation in a small set of fractal dimension less than three, the inverse cascade of two-dimensional turbulence can also display lack of self-similarity and intermittency if the energy injection is constrained in a fractal set of dimension less than two. 
A series of numerical simulations  of two dimensional turbulence are examined, using different forcing functions of the same forcing length-scale but different fractal dimension $D$ that varies from the classical $D=2$ case to the point vortex case $D=0$. It is shown that as the fractal dimension of the forcing is decreased from $D=2$, the
self-similarity is lost and intermittency appears, with the scaling of the different structure functions $\langle |\delta u_\|^p| \rangle\propto r^{\zeta_p}$ differs from the dimensional analysis prediction $\zeta_p=p/3$. The present model thus provides a unique example that intermittency is controlled and can thus shed light and provide test beds for multi-fractal models of turbulence. 
\end{abstract}

\maketitle


\section{\label{sec:level1}  Introduction }

Turbulence is pervasive in natural and industrial flows.
In his first statistical description of turbulence Kolmogorov \cite{K41}  argued that energy in turbulent flows cascades to smaller and smaller scales in such a way that there is a constant flux of energy from the large scales where energy is injected to the small viscous scales where energy is dissipated. Assuming further that this process is self-similar lead to the prediction that the different moments of velocity differences 
\beq 
S_p(r) \equiv  \left\langle \left|\frac{\bf r }{r} \cdot(\bu(\bx+\br)-\bu(\bx) )\right|^p\right\rangle \label{sp}
\eeq
separated by a distance $r=|{\bf r}|$ scale like 
$S_p\propto r^{p/3}$ with the case $p=3$ being an exact result (without the absolute value in eq.\ref{sp}).
There is a mass of evidence however from the past years that this result is not exact; self-similarity is broken and the powers of velocity differences scale with different exponents $S_p(r) \propto r^{\zeta_p}$ where $\zeta_p\ne p/3$. This breaking of self-similarity is referred to as intermittency. It appears because as the cascade develops towards smaller scales, energy is concentrated in a set that occupies a smaller and smaller fraction of the domain volume 
so that finally energy dissipation is concentrated
in a fractal set of dimension smaller than three
\cite{frisch1995turbulence,alexakis2018cascades}. Modern theory of turbulence attempts to understand quantitatively the origin of intermittency and predict these exponents.

In two dimensions on the other hand, due to the presence of a second invariant, enstrophy, the energy cascades in an inverse way from small to large scales \cite{boffetta2012two}. This behavior was first predicted by Kraichnan–Leith–Batchelor (KLB) theory \cite{batchelor1969computation,kraichnan1967inertial,leith1968diffusion}. What was equally interesting was that the inverse cascade of energy in two dimensions is in fact
self-similar so that all moments of velocity differences scale with $r$ with exponents $\zeta_p=p/3$ \cite{boffetta2000inverse}. 
This is explained by the fact that larger eddies extract energy from an ensemble of smaller eddies averaging out this way any extreme events.
As energy moves up in scale it is not restricted in a set of dimension other than two.
This behavior however does not always have to be the case as we argue in this work. If the energy injection in two dimensional turbulence is not space filling but is restricted in a set of dimension $D$ smaller than two then as energy moves up in scale it can occupy larger and larger area fraction so that only at the largest scale it is concentrated in a two-dimensional set.

Fractal forcing is not just an abstract construction. In engineering it has been employed extensively in three dimensional turbulence with the use of fractal grids in simulations and experiments in order to enhance turbulence \cite{hurst2007scalings,nagata2013turbulence,seoud2007dissipation,laizet2011dns,mazzi2004fractal}
but not as far as we know in two dimensions.
In nature, atmospheric and oceanic flows are close to two-dimensional. 
When driven by winds over rough topography \cite{bretherton1976two,vallis1993generation} resemble two-dimensional turbulence driven by a fractal forcing.
Furthermore, quasi-two-dimensional flows are believed to transition to an inverse cascade in a critical manner \cite{alexakis2018cascades}. 
In such flows the energy injected in the two-dimensional manifold appears in a set of smaller dimension occupying a fraction of the domain area that approaches zero as criticality is approached \cite{benavides2017critical,seshasayanan2014edge}.

In this work we show using an extensive set of numerical simulations that indeed intermittency 
can appear in the inverse cascade of energy when the energy-injection mechanism is restricted to a set of fractal dimension $D<2$. This model does not only give new insights in two-dimensional turbulence but also provides a unique example that intermittency is controlled and can thus provide test beds for multi-fractal models of turbulence.

\section{Numerical simulations} 
We begin by considering  the incompressible flow
in a double periodic square domain of side $2\pi L$.
In terms of the vorticity  $\omega$ the two dimensional Navier-Stokes equation can be written as
\beq 
    \partial_t \omega + \bu \cdot \nabla \omega =
    \nu \nabla^2 \omega + \alpha \nabla^{-2}\omega + f_\omega
\eeq 
where  
the velocity $\bu$ is linked to $\omega$ by $\omega=\curl \bu$, 
$\nu$ is the viscosity and $\alpha$ is a hypo-viscosity used to absorbs energy arriving at the largest scales at a rate $\epsilon_\alpha=\langle |\nabla^{-1} \bu|^2\rangle$. The curl of the forcing is given by $f_\omega$ that injects energy at a rate $\epsilon$ at a lengthscale $\ell_f$. Given $f_w$ there are three independent non-dimensional control numbers: the Reynolds number $Re=\epsilon^{1/3}\ell_f^{4/3}/\nu$, the hypo-viscous Reynolds number $Re_\alpha=\epsilon^{1/3}\ell_f^{-8/3}/\alpha$ and the domain to forcing scale ratio $\lambda =L/\ell_f$. This system of equations was solved numerically using the pseudo-spectral code {\sc ghost} \cite{mininni2011hybrid} with 2/3 de-aliasing  and second order Runge-Kutta 
method for the time advancement. Since we are interested in the inverse cascade the Reynolds numbers was kept fixed to small value
$Re=10$. This value of $Re$ is sufficiently large to allow for the development of the inverse cascade but suppresses the forward enstrophy cascade
and any intermittency related to it. As a result the smallest scales in the system are given by $\ell_f$ and energy in these scales is concentrated close to the forcing. The hypo-viscous Reynolds number $Re_\alpha$ was set to $Re_\alpha=10 \lambda^{8/3}$ so that the large scale dissipation lengthscale $\ell_\alpha$ remains fixed and close to the domain size $\ell_\alpha\simeq L$. Five different resolutions $N$ were used varying $\lambda$ as given in the table \ref{table}. 
\begin{table}
\begin{center}
\begin{tabular}{|c| c c c c c c c| }
\hline
$N$              & 512  & & 1024 & & 2048  & & 4096  \\  \hline
$\lambda$        & 16  &  & 32  &  &  64   & &  128  \\ 
$ Re_\alpha$     & $2.2\cdot 10^4$ &$ \quad$& $1.4 \cdot 10^5$& $ \quad$& $8.9\cdot 10^5$ &$ \quad$ & $5.6 \cdot 10^6$ \\ 
 \hline
\end{tabular}
\caption{\label{table} Resolution $N$, scale separation $\lambda=L/\ell_f$
and hypo-viscous Reynolds number $Re_\alpha$.}
\end{center}
\end{table}

\begin{figure}
\includegraphics[width=0.49\textwidth]{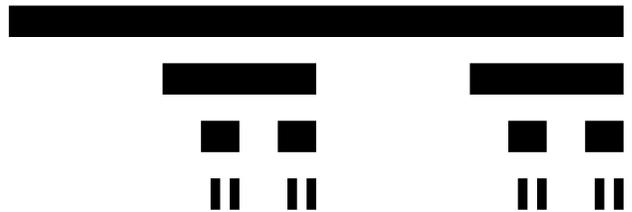}
\caption{\label{fig:fractal} Demonstration of how a fractal set of dimension 1/2 is formed. At every step (down) the initial set is split 
in 4 equal sub-sets out of which subset 1 and 3 are disregarded. }
\end{figure}

\begin{figure*} 
\includegraphics[width=0.99\textwidth]{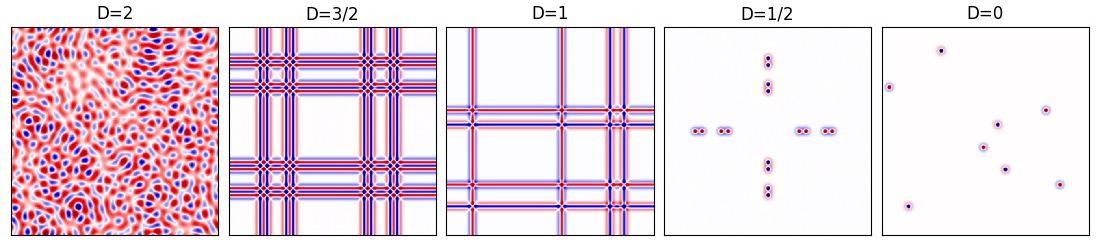}
\includegraphics[width=0.99\textwidth]{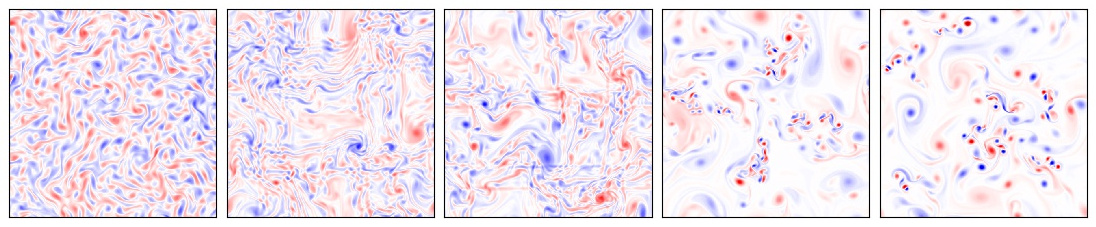}
\caption{\label{fig:Vorticity} Top panels: Random forcing function for different dimension $D$ at resolution $N=512$. Lower panels: Vorticity of the flow for the same cases as above at steady state.}
\end{figure*}

Finally, five different forcing functions of different fractal dimension $D$ are considered.  The first one corresponding to $D=2$ is the classical  random forcing where all Fourier modes of wave-vectors $\bk$ satisfying  $|{\bf k}|\simeq 1/\ell_f$ are forced with random phases. The $D=1$ forcing corresponds to four vertical and four horizontal vortex lines with Gaussian profiles of width $\ell_f$ randomly placed  in the domain. Similarly $D=0$ corresponds to eight point-vortexes with Gaussian profile of width $\ell_f$ randomly placed in the domain.
The $D=3/2$ and $D=1/2$ correspond to Cantor sets  that are constructed as follows. 
For $D=3/2$ a dense set of horizontal and vertical vortex-lines are uniformly placed in the domain. This set is split in four equal subsets from which subset one and three are removed. The remaining sets are then split again in four from which again subset one and three are removed and so on, as demonstrated in figure \ref{fig:fractal}, until no further splitting can be done. The resulting box-counting dimension is $D=3/2$ \cite{triebel2010fractals}. For $D=1/2$  we start with point-vortexes placed along one vertical and one horizontal line and we follow the same procedure  leading this time to a box-counting dimension $D=1/2$. For all forcing functions the the forcing lengthscale $\ell_f$ was fixed so that the peek of the forcing spectrum was around similar wavenumber $k_f\simeq 1/\ell_f$. Furthermore, in all cases the amplitude of the forcing function was varied randomly delta-correlated in time fixing thus the energy injection rate $\epsilon$. A color plot of the forcing functions for the five forcing functions is shown in the upper panels of figures \ref{fig:Vorticity}. For this figure we used the smallest $\lambda$ (largest $\ell_f$) so that the point-vortexes in the left panel are clearly visible. 
\begin{figure}[b]
\includegraphics[width=0.49\textwidth]{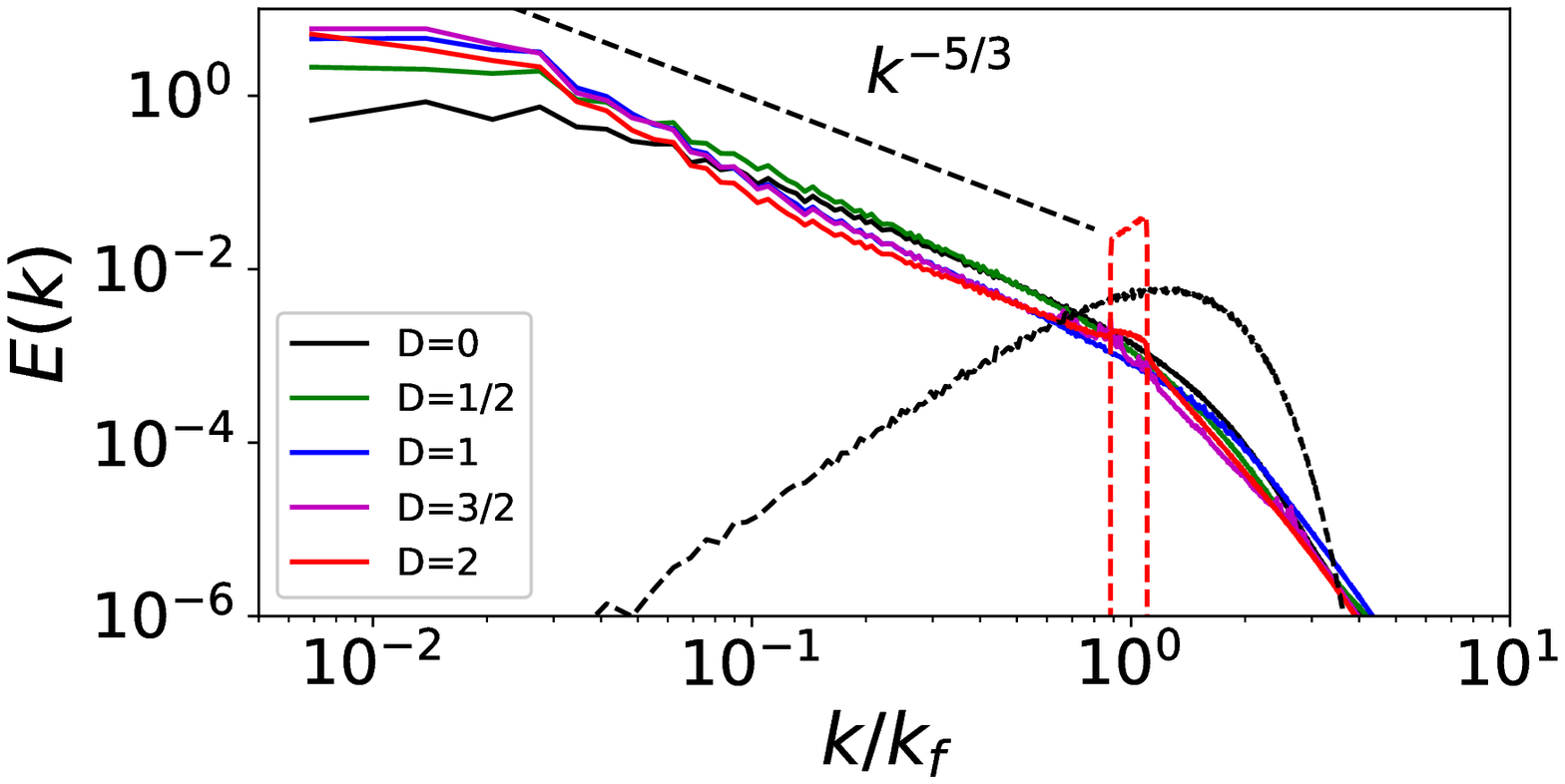}
\includegraphics[width=0.49\textwidth]{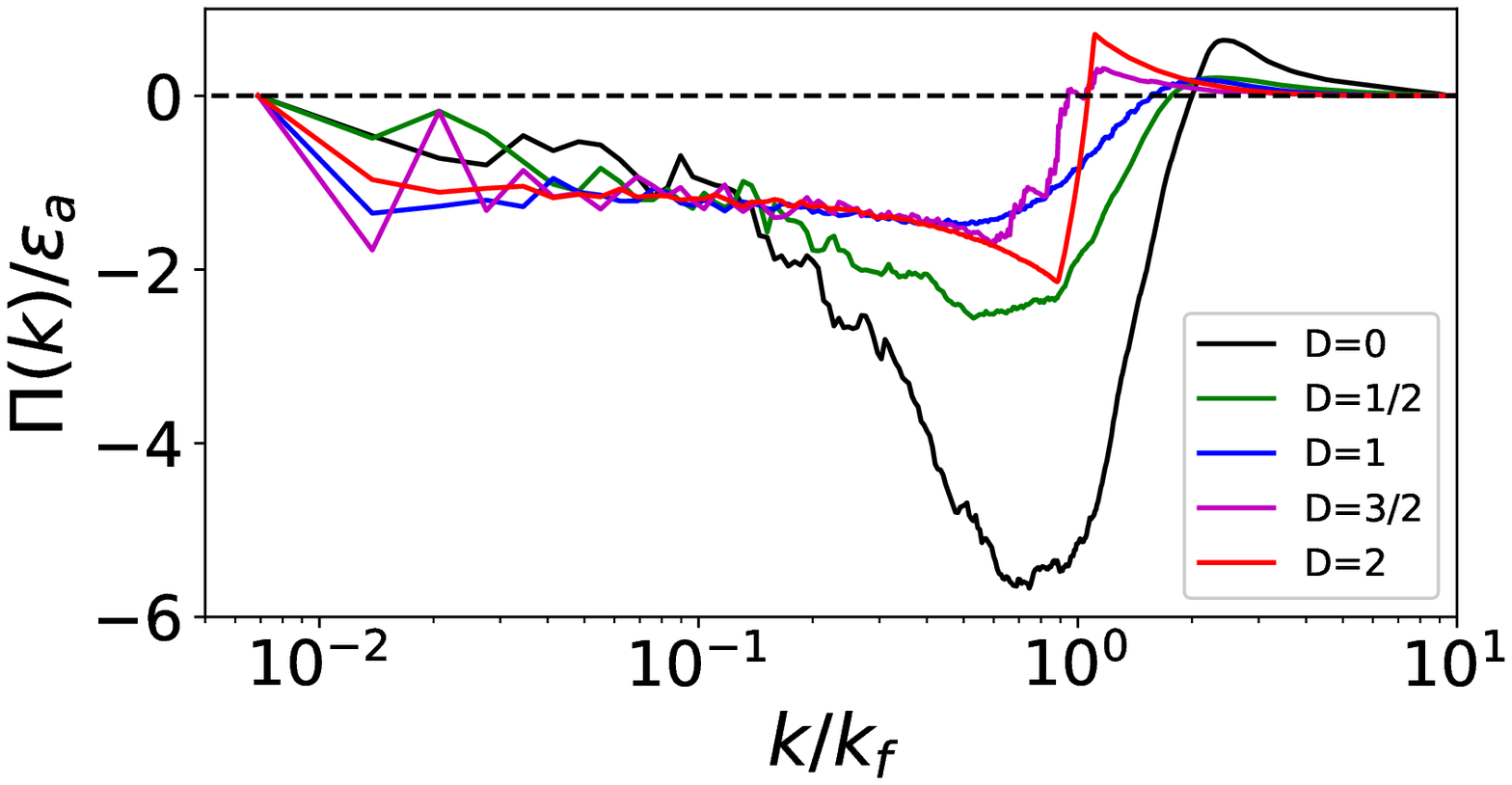}
\caption{\label{fig:spectra} Top: Energy spectra for the highest resolution runs. The dashed lines give the forcing spectrum for the two extreme cases $D=2$ and $D=0$. The straight dashed line gives the $k^{-5/3}$ scaling.
Bottom: Energy flux for the same cases. 
}
\end{figure}

\section{Results}
All forcing functions lead to an inverse cascade of energy 
marked by an energy spectrum close to $k^{-5/3}$ (top panel of figure \ref{fig:spectra})
and a negative flux of energy $\Pi(k)=\langle \bu^<_k \cdot (\bu \cdot \nabla \bu)\rangle$
(where $\bu^<_k$ indicates the field $\bu$ filtered so that only wavenumbers with $|\bk|\le k$ are kept)
shown in the lower panel of figure \ref{fig:spectra}. 
The flux is approximately constant for the $D\ge1$ cases but due to the relative small $Re$ and the fact that the peek of the forcing
spectrum for the cases $D=3/2$ and $D=2$ is slightly higher, the flux is affected by finite $Re$ effects displaying a decrease close to the forcing scales.   
\begin{figure}
\includegraphics[width=0.49\textwidth]{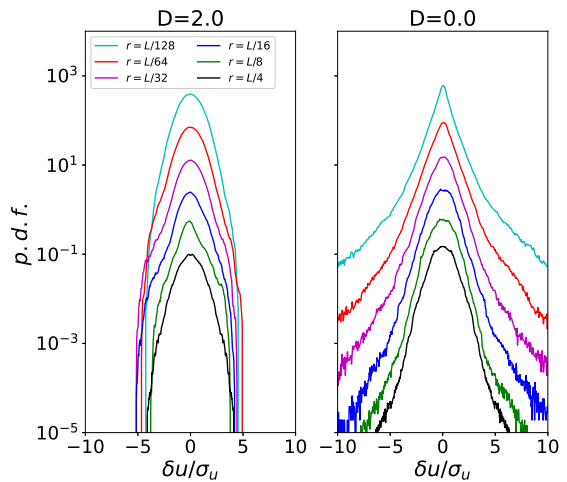}
\caption{ The probability distribution function of $\delta u(r)$ 
normalized by its variance $\sigma_u=S_2(r)^{1/2}$ for different values of $r$ and for $D=2$ left
and $D=0$ right. The curves have been shifted vertically for clarity.
\label{fig:pdf} }
\end{figure}
\begin{figure}
\includegraphics[width=0.49\textwidth]{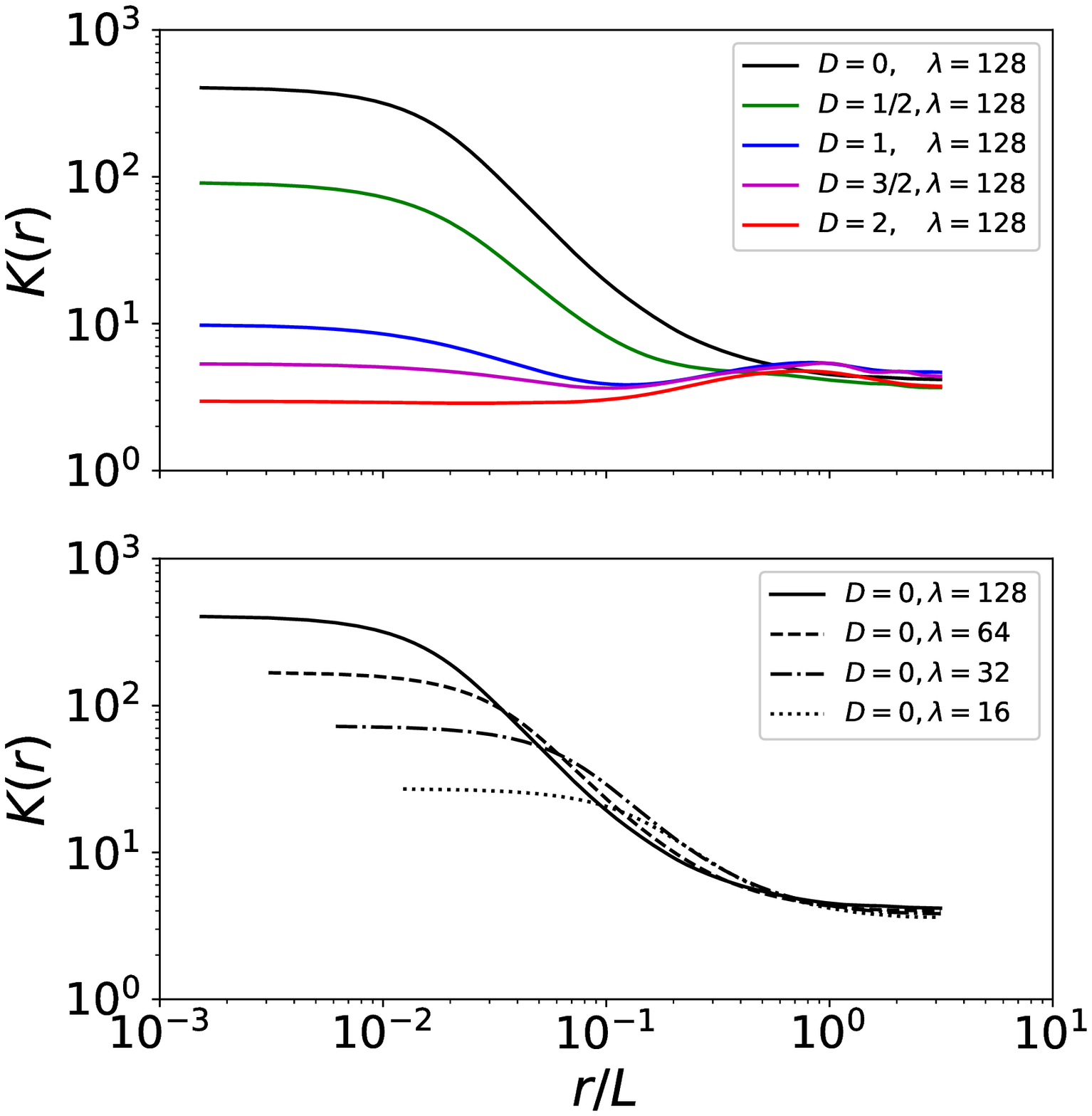}
\caption{The top panel shows $K(r)$ for the flows at highest $\lambda$ for the five different forcing functions. The lower panel shows $K(r)$ for $D=0$
and four different values of $\lambda$.
\label{fig:Kurtosis} }
\end{figure}

Although all the five cases have an inverse cascade they
do not have the same turbulent statistical behavior. This can already be seen in the vorticity plots shown in the lower panels of figure \ref{fig:Vorticity}. Turbulence, marked by intense vorticity regions, is uniformly spread in the domain for the $D=2$ case but as 
$D$ is decreased intense vorticity regions occupy a smaller area fraction but with larger intensity. 
In order to quantify this observation we plot in figure \ref{fig:pdf} the probability distribution function (p.d.f.) of velocity differences 
for the two extreme cases $D=2$ and $D=0$ for different values of $r$ starting from the largest $r=L/4$ to the forcing scale $r=L/128=\ell_f$. 
In the $D=2$ case the p.d.f.s are close to Gaussian for all examined $r$.
Furthermore, no significant change is observed as $r$ is varied: ie they are self-similar.
In the $D=0$ case on the other hand the p.d.f.s deviate from the Gaussian distribution having large tails. 
Most importantly, as smaller values of $r$ are considered the deviations from Gaussianity become stronger
with the distribution becoming more peeked and with stronger tails. In other words self-similarity is lost 
for the $D=0$ case.


A quantitative way to measure this lack of self-similarity,  is to measure the kurtosis 
$K(r)=S_4(r)/S_2^2(r)$.
Kurtosis gives a measure of how heavy-tailed is the distribution of $\delta u$.
$K(r)=3$ corresponds to a Gaussian distributed field while larger values correspond 
to fields of wider distribution. 
If the distribution is self-similar $K(r)$ will be independent of $r$. 
In figure \ref{fig:Kurtosis} we plot the Kurtosis for different cases. 
The top panel shows $K(r)$ for the flows at highest $\lambda=128$ (highest resolution) for the five different forcing functions. 
For $D=2$, $K(r)$ is almost flat and close to $3$, indicating that $\delta u$ follows a self-similar, nearly-Gaussian
distribution. As the dimension of the forcing is decreased $K(r)$ takes larger and larger values in the small $r$ range,
with the $D=0$ case having two orders of magnitude larger $K(\ell_f)$ than a Gaussian field.
The lower panel shows the case $D=0$ for the different values of $\lambda$. 
As $\lambda$ and $Re_\alpha$ are increased the non-self-similar behavior extends to a 
larger range of $r$ with the deviation from Gaussianity increasing. Therefore this amounts to a phenomenon that persists 
and extends as the infinite $Re_\alpha$ and infinite box-size limit is reached.

\begin{figure}
\includegraphics[width=0.49\textwidth]{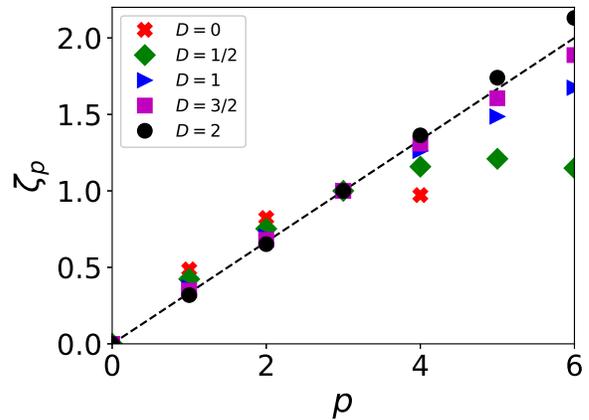}
\caption{\label{fig:ZETA} Scaling exponents $\zeta_p$ for the five different $D$ considered calculated using extended-self-similarity.}
\end{figure}
The result of figure \ref{fig:Kurtosis} already indicates that $\zeta_2$ and $\zeta_4$ can not follow the scaling $\zeta_p=p/3$
that would imply an $r$ independent Kurtosis. In figure \ref{fig:ZETA} we plot the exponents $\zeta_p$ for all cases.
The exponents were measured using the extended self-similarity assumption \cite{benzi1993extended}  to extend the range that a power-law behavior is observed
where $S_p(r)$ is plotted as a function of $S_3(r)$ and assuming the theoretically predicted linear scaling of the third moment $S_3(r)\propto r$.  
The exponents up to $p=6$ were calculated except for the $D=0$ case that showed a particular slow convergence for the high $p$ moments.
For $D=2$ the exponents follow the linear scaling $\zeta_p=p/3$. As $D$ is decreased, the exponents with $p<3$ increase while exponents with $p>3$ decrease. The flow thus becomes more intermittent as $D$ is decreased with the high order moments being dominated by few but strong events in the tail of the distribution.  
 
\section{Conclusions}

In this work we have shown, that the inverse cascade of energy can display intermittent features provided that the forcing function injects energy in fractal a set of dimension smaller than two. Intermittency is demonstrated by long tails in the distribution of the velocity differences at small scales caused by the forcing, that as larger scales are approached they flatten out becoming closer to Gaussian. This behavior was shown to persist as larger domains (larger $\lambda$) are considered and become stronger as the dimension $D$ is decreased.

Here we employed only mono-fractal forcing. Bi-fractal, or multi-fractal forcing can also be considered by adding with appropriate weight different fractal forcing functions. This could lead to an intermittent behavior that is closer to the three dimensional cascade, (albeit in the opposite direction). 
More generalized fractal forcing functions that dynamically evolve could also be considered. Such functions could be examined in future work.

Most importantly the strength of intermittency caused by the fractal forcing is determined by its dimension and thus it provides a way to create a cascade for which the intermittency can be varied from self-similar to strongly intermittent.  Ideas, about the behavior and origin of intermittency can thus be put in the test using this simple model.  Finally one can ask how a fractal forcing can affect intermittency properties in three dimensional turbulence.

\begin{acknowledgments}
This work was granted access to the HPC resources of MesoPSL financed by the Région Île-de-France and the project EquipMeso (project no. ANR-10-EQPX-29-01), of the HPC resources of GENCI-TGCC \& GENCI-IDRIS (project no. A0110506421) and the  HPC facility ARIS from the Greek Research and Technology Network (GRNET).
AA is supported by the project Dysturb (project no. ANR-17-CE30-0004) finnanced by the Agence Nationale pour la Recherche (ANR).
\end{acknowledgments}

\bibliography{2DIntermittency}

\end{document}